\documentclass[iopart,twocolumn,superscriptaddress]{revtex4-2}
\usepackage{amssymb}
\usepackage{amsmath}
\usepackage{mathrsfs}
\usepackage{graphicx}
\usepackage{float}
\usepackage[colorlinks=True,citecolor=blue,linkcolor=blue,allcolors=blue]{hyperref}
\usepackage[normalem]{ulem}
\usepackage{color}
\usepackage{bm}

\begin{document}

\title{\textcolor{black}{Multiple topological transitions and spectral singularities in non-Hermitian Floquet systems}}

\author{Weiwei Zhu}
\email{phyzhuw@ouc.edu.cn}
\affiliation{College of Physics and Optoelectronic Engineering, Ocean University of China, Qingdao 266100, China}
\affiliation{Engineering Research Center of Advanced Marine Physical Instruments and Equipment of Education Ministry, Ocean University of China, Qingdao 266100, China}
\affiliation{Qingdao Key Laboratory for Optics Photoelectronics, Ocean University of China, Qingdao 266100, China}
\author{Longwen Zhou}
\email{zhoulw13@u.nus.edu}
\affiliation{College of Physics and Optoelectronic Engineering, Ocean University of China, Qingdao 266100, China}
\affiliation{Engineering Research Center of Advanced Marine Physical Instruments and Equipment of Education Ministry, Ocean University of China, Qingdao 266100, China}
\affiliation{Qingdao Key Laboratory for Optics Photoelectronics, Ocean University of China, Qingdao 266100, China}
\author{Linhu Li}
\affiliation{Guangdong Provincial Key Laboratory of Quantum Metrology and Sensing $\&$ School of Physics and Astronomy, Sun Yat-Sen University (Zhuhai Campus), Zhuhai 519082, China}
\author{Jiangbin Gong}
\email{phygj@nus.edu.sg}
\affiliation{Department of Physics, National University of Singapore, 117551, Singapore}
\affiliation{Centre for Quantum Technologies, National University of Singapore, 117543, Singapore}

\begin{abstract}
\textcolor{black}{The interplay between Floquet driving and non-Hermitian gain/loss could give rise to intriguing phenomena including topological funneling of light, edge-state delocalization, anomalous topological transitions and Floquet non-Hermitian skin effects. In this work, we uncover two unique phenomena in Floquet systems caused by gain and loss. First, multiple topological transitions from anomalous Floquet second-order topological insulators to anomalous Floquet first-order topological insulators and then to normal insulators can be induced by gain and loss. Interestingly, the resulting anomalous Floquet insulators further carry hybrid skin-topological boundary modes, which could either be fully localized or localized to different edges at different time slices and traversing along all edges in a single driving period.  The topological phase transitions are also shown to be detectable through studies of transmission properties in the setting of  coupled ring resonators.    Second, gain and loss are found to induce singularities in the Floquet spectral, around which anomalous transmissions at flat quasienergy bands are predicted. These discoveries not only enhanced our understanding of topological matter and phase transitions in driven non-Hermitian systems, but also promoted their experimental realizations in optical and acoustic settings.}
\end{abstract}

\maketitle
\section{Introduction}\label{sec:Int}

\textcolor{black}{Floquet systems, whose Hamiltonians are time-periodic, have attracted great attention over the past decades and played important roles in the study of quantum dynamics under external driving fields. Counterintuitive phenomena, such as dynamical localization~\cite{PhysRevLett.67.516,PhysRevLett.69.1986,PhysRevLett.91.110404,PhysRevLett.117.144104,NC7653} and stabilization~\cite{PhysRevLett.65.2362,PhysRevA.43.2474,PhysRevLett.73.1777,doi:10.1126/science.262.5137.1229,BUCHLEITNER2002409,Gong2015} have been revealed in Floquet systems. More recently, time-periodic drivings have been applied to engineer Floquet topological matter
~\cite{PhysRevB.79.081406,NatPhys7.490,PhysRevB.84.235108,Gong2012}. Irradiated by electromagnetic fields, semimetals and normal insulators could be driven into topological nontrivial phases
~\cite{PhysRevB.84.235108,NatPhys16.38,PhysRevLett.121.036401,NatCommun8.13940}, which may support novel topological states that are absent in static systems, such as Floquet winding metals~\cite{ZhouPRB2016,PhysRevLett.118.105302,PhysRevLett.130.056901}, first-order
~\cite{PhysRevX.3.031005,PhysRevLett.117.013902,NatCommun7.13368,NatCommun8.13756,NatPhys16.1058,ZhouPRB2018} and higher-order~\cite{PhysRevLett.124.216601,PhysRevB.103.L041402,PhysRevB.104.L020302,NatCommun13.11,PhysRevB.105.115418,PhysRevB.104.L140502,ZhouPRB2019} anomalous Floquet topological insulators. These nonequilibrium topological phases have also been widely explored in photonic, acoustic and cold atom experiments~\cite{NatCommun7.13368,NatCommun8.13756,NatPhys16.1058,NatCommun13.11,Nature496.196,PhysRevLett.130.056901,NatMater21.634,LSA9.128}.}

\textcolor{black}{Non-Hermitian effects, which are ubiquitous in open systems, have been recently employed as efficient means to actively control the physical properties of photonic, acoustic and cold atom setups~\cite{NatPhys14.11,annurev,NRP4.745,Wang_2021,RevModPhys.93.015005}. With complex energies, non-Hermitian systems may support richer topological structures that originate from the windings of their eigenspectra~\cite{annurev,NRP4.745,Wang_2021,RevModPhys.93.015005,Rijia53605}, rather than the geometric phase of eigenfunctions like in Hermitian systems. Exceptional points (EPs)~\cite{PhysRevLett.86.787,NM18,science.aar7709,Nature537} and non-Hermitian skin effects (NHSEs)~\cite{PhysRevLett.121.086803,PhysRevB.99.201103,PhysRevLett.124.086801,NC11} are two representative examples that exhibit topological features unique to non-Hermitian systems.}


\textcolor{black}{The marriage between Floquet engineering and non-Hermitian control forms a natural extension in the study of driven open systems. In recent years, interesting discoveries have been made on non-Hermitian Floquet topological matter~\cite{PhysRevB.98.205417,Xi2020,PhysRevLett.123.190403,PhysRevB.101.014306,PhysRevResearch.3.023211,PhysRevB.102.041119,Nature601,sciadv.abo6220,e25101401}, including Floquet EPs and Floquet NHSEs. In experiments, gain and loss can be utilized as efficient means to achieve the non-Hermitian control of Floquet systems. Topological funneling of light~\cite{science.aaz8727}, delocalized topological edge modes~\cite{PhysRevB.103.195414}, Floquet NHSEs~\cite{ZhouPRB2021,PhysRevB.106.134112}, non-Hermitian topological phase transitions~\cite{PhysRevLett.125.013902,NP20}, higher-order NHSEs~\cite{NC12}, Floquet hybrid NHSEs~\cite{PhysRevB.106.035425,PhysRevB.108.L220301,PhysRevLett.132.063804} and transient NHSEs~\cite{NaC13} have been theoretically proposed or experimentally realized along this line of thought. Considering the rich physics of Floquet and non-Hermitian systems, the interplay between Floquet engineering and gain/loss effects should bring about intriguing new phenomena that are awaited to be further unveiled.}


\textcolor{black}{In this paper, we uncover two unique phenomena not reported before in non-Hermitian Floquet systems, i.e., gain/loss-induced multiple topological transitions and anomalous wave transmissions due to the singularity of Floquet spectrum. We introduce spatially modulated gain and loss into a Floquet bipartite lattice, which supports anomalous Floquet second-order topological insulators (AFSOTI) under suitably chosen parameters~\cite{PhysRevB.103.L041402}. With the increase of gain/loss strengths, we identify two topological transition points, across which the gaps between Floquet bands close and reopen. The first point is associated with a transition from AFSOTI to anomalous Floquet topological insulators (AFTI). The second point corresponds to a transition from AFTI to normal insulators (NI). Due to the existence of gain and loss, the AFTI also possesses hybrid-skin topological effects~\cite{PhysRevLett.123.016805}, and the boundary modes of such states are found to be geometry-dependent. For a horizontal square, the boundary modes are localized at the left-down corner for higher-order skin effects. For a square at $45^{\circ}$ to the horizontal line, the boundary modes are instead localized to different edges at different time slices and traversing all edges in one driving period. Such geometry-dependent effect can be understood by tracing the evolution of topological edge modes in coupled ring resonators. We further study the scattering properties of the system by the transfer-matrix method, and discover anomalous transmissions in certain parameter regions, even though the Floquet bands of the system are flat. These anomalous transmissions are found to originate from the Floquet-spectral singularities, whose locations in the parameter space are further sensitive to the system size.}

\textcolor{black}{This paper is organized as follows. In Sec.~\ref{sec:Mod}, we introduce our model and outline its theoretical descriptions including the tight-binding approach and the transfer matrix method. In Sec.~\ref{sec:Res}, we reveal the multiple topological transitions through examining the changes in the Floquet bands and provide topological characterizations for different existing phases. These results are verified by quasienergy spectrum calculations under different boundary conditions. The geometry-dependent effects of gain/loss-induced AFTI are further discussed. In Sec.~\ref{sec:SS}, we unveil the transmission properties of our system, which are consistent with quasienergy-band calculations in most cases. However, in a small parameter region where the quasienergy band is flat, the transmission is found to be unexpectedly large, which is due to the singularities in the Floquet spectrum. In Sec.~\ref{sec:Sum}, we conclude our study, discuss potential applications of the corner skin modes and consider their possible experimental realizations.}


\section{Model}\label{sec:Mod}

\begin{figure}
\includegraphics[width=0.9\linewidth]{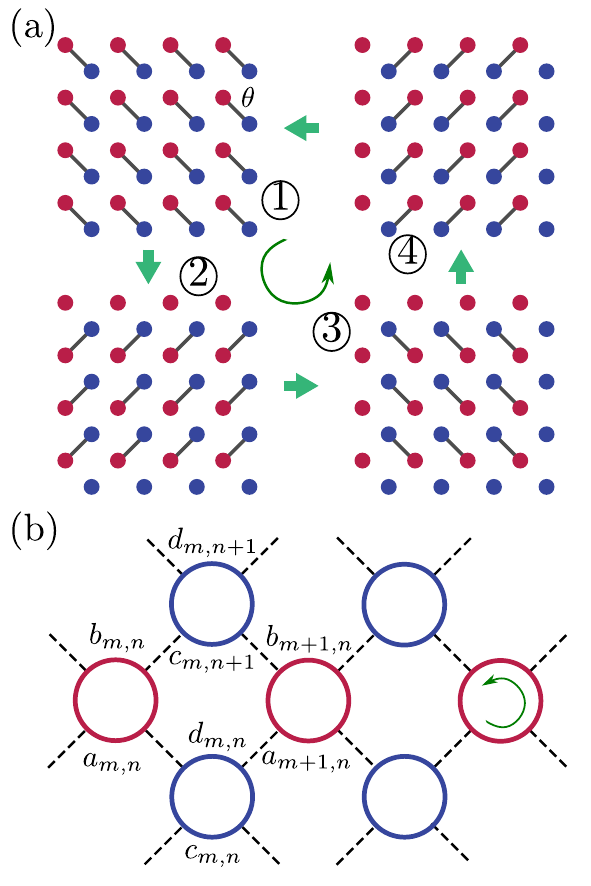}
\caption{Models of two-dimensional $\mathcal{PT}$-symmetric Floquet bipartite lattices. (a) The model under a four-step driving protocol. The couplings in each step are fully dimerized.  Each site is sequentially coupled with one of the four neighboring sites in a counterclockwise direction. Gain and loss are alternatively distributed in space. Red (blue) sites represent gain (loss) sites. Grey lines represent nearest-neighbor couplings. (b) The equivalent coupled ring resonators model. Dashed lines denote couplings between nearest-neighbor resonators.}
\label{model}
\end{figure}

\textcolor{black}{We start with a time-dependent, tight-binding lattice model, whose driving protocol is illustrated in Fig.~\ref{model}(a). One driving period contains four steps, with dimerized intersite couplings introduced in each step. Each site is sequentially coupled with one of its four neighboring sites in a counterclockwise direction. Every unit cell has two sublattices with balanced gain and loss. Under the periodic boundary condition, the time-dependent, four-step Hamiltonian in momentum space is given by}

\begin{eqnarray}
 H(\mathbf{k},t)=\left\{
\begin{array}{cc}
H_{1}(\mathbf{k})&\ell T<t\leq \ell T+T/4\;,\\
H_2(\mathbf{k})&\ell T+T/4<t\leq \ell T+T/2\;,\\
H_3(\mathbf{k})&\ell T+T/2<t\leq \ell T+3T/4\;,\\
H_4(\mathbf{k})&\ell T+3T/4<t\leq \ell T+T\;,
\end{array}
\right.
\label{eq1}
\end{eqnarray}
where $\ell\in\mathbb{Z}$ and 
\begin{equation}
H_{m}(\mathbf{k})=\theta(e^{i{\bf b}_m\cdot\mathbf{k}}\sigma^{+}+{\rm h.c.})+ig\sigma_z
\label{eq2}
\end{equation}
for $m=1,2,3,4$. Here, $\theta$ is a coupling parameter, $g$ is the strength of gain and loss, $\sigma_{x,y,z}$ are Pauli matrices, and $\sigma^{\pm}=(\sigma_{x}\pm i\sigma_{y})/2$. $T$ is the driving period and we set $T=4$ in dimensionless units. The vectors $\mathbf{b}_{m}$ are given by $\mathbf{b}_{1}=(0,0)$, $\mathbf{b}_{2}=(0,1)$, $\mathbf{b}_{3}=(-1,1)$ and $\mathbf{b}_{4}=(-1,0)$. The Hamiltonian possesses $\mathcal{PT}$ symmetry $\mathcal{PT}H(\mathbf{k},t)(\mathcal{PT})^{-1}=H(\mathbf{k},t)$, non-Hermitian particle-hole symmetry (NHPHS) $\mathcal{CK}H(\mathbf{k},t)(\mathcal{CK})^{-1}=-H(-\mathbf{k},t)$, and pseudo-inversion symmetry (PIS) $\mathcal{I}H(\mathbf{k},t)\mathcal{I}^{-1}=H^{\dag}(-\mathbf{k},t)$. Here $\mathcal{P}=\mathcal{I}=\sigma_x$, $\mathcal{C}=\sigma_z$, $K$ represents complex conjugate and $\mathcal{T}=K$ is the (spinless) time-reversal operator.

The quasienergy dispersion $\varepsilon({\bf k})$ can be obtained by solving the Floquet eigenvalue equation,
\begin{equation}\label{eq3}
  U_T(\mathbf{k})\, |\Psi(\mathbf{k})\rangle = e^{-i\varepsilon(\mathbf{k})} |\Psi(\mathbf{k})\rangle\;,
\end{equation}
where $U_T(\mathbf{k}) \equiv \mathfrak{T}\mathrm{exp}\big[\!-\!i\!\int_{0}^{T} H(\mathbf{k},\tau)\,d\tau\big]$ is the Floquet operator. $\mathfrak{T}$ is the time-ordering operator. 
\textcolor{black}{In the $\mathcal{PT}$-invariant region, the quasienergy bands are real. The $\mathcal{PT}$ symmetry further enforces the quasienergy bands to satisfy $\varepsilon_n(\mathbf{k})=\varepsilon^*_m(\mathbf{k})$ for $n\neq m$, where $n$ and $m$ are band indices. In the $\mathcal{PT}$-broken region, the Floquet bands thus form complex conjugate pairs. Besides, the Floquet bands are also constrained by the NHPHS [PIS] and satisfy $\varepsilon_n(\mathbf{k})=-\varepsilon^*_m(-\mathbf{k})$ [$\varepsilon_n(\mathbf{k})=\varepsilon^*_m(-\mathbf{k})$]. The PIS requires the band inversion to happen at high-symmetry momentum points ($\Gamma$ points in our model), where topological phase transitions happen.}

The four-step Hamiltonian in Eq.~(\ref{eq1}) is equivalent to coupled ring resonators array~\cite{PhysRevB.89.075113,PhysRevLett.110.203904,NaC7}, as illustrated in Fig.~\ref{model}(b). In each ring resonators, only anticlockwise modes are considered. Red (blue) ring resonators correspond to gain (loss) sites. Each gain (loss) ring resonators is only coupled with its nearest-neighbor resonators. The transport properties of coupled ring resonators array can be described by the transfer matrix method (under periodic boundary condition along $y$ with $a_{m,n}=a_m e^{ik_yn}$, $b_{m,n}=b_m e^{ik_yn}$, $c_{m,n}=c_m e^{ik_yn}$, $d_{m,n}=d_m e^{ik_yn}$), i.e.,
\begin{equation}\label{eq4}
  \left(\begin{array}{c}
     a_{m+1} \\
     b_{m+1}
   \end{array}
  \right)=M'\left(\begin{array}{c}
                  c_m \\
                  d_m
                \end{array}
  \right)=M'M\left(\begin{array}{c}
                       a_m \\
                       b_m
                     \end{array}
  \right),
\end{equation}
where $M$ and $M'$ are two by two matrices. They can be constructed from the scattering matrices $S$ and $S'$ by
\begin{eqnarray}
  M_{11} =\frac{S_{12}S_{21}-S_{11}S_{22}}{S_{12}},\ \ M_{12} =\frac{S_{22}}{S_{12}}, \\
  M_{21} =-\frac{S_{11}}{S_{12}},\ \ \ \ \ \ \ \ \ \ \ \ \ \ \ M_{22} = \frac{1}{S_{12}}, \\
  M'_{11} =\frac{S'_{12}S'_{21}-S'_{11}S'_{22}}{S'_{21}},\ \ M'_{12} =\frac{S'_{11}}{S'_{21}}, \\
  M'_{21} =-\frac{S'_{22}}{S'_{21}},\ \ \ \ \ \ \ \ \ \ \ \ \ \ \ M'_{22} = \frac{1}{S'_{21}},
\end{eqnarray}
with
\begin{alignat}{1}
	S= & \begin{pmatrix}e^{i\varepsilon/4} & 0\\
		0 & e^{i(\varepsilon/4-k_{y})}
	\end{pmatrix}s\begin{pmatrix}e^{i\varepsilon/4} & 0\\
		0 & e^{i(\varepsilon/4+k_{y})}
	\end{pmatrix}s,\\
	S'= & s\begin{pmatrix}e^{i\varepsilon/4} & 0\\
		0 & e^{i(\varepsilon/4-k_{y})}
	\end{pmatrix}s\begin{pmatrix}e^{i\varepsilon/4} & 0\\
		0 & e^{i(\varepsilon/4+k_{y})}
	\end{pmatrix},
\end{alignat}
where $(b_m, c_m)^T=S(a_m, d_m)^T$, $(a_{m+1}, d_m)^T =S'(b_{m+1}, c_m)^T$, and $s\equiv e^{i(\theta\sigma_x+ig\sigma_z)}$.

\section{Gain/loss-induced multiple topological phase transitions}\label{sec:Res}

\subsection{Non-Hermitian topological phase transitions}\label{subsec:NHTPT}
\textcolor{black}{In the absence of gain and loss, the time-dependent Hamiltonian in Eq.~(\ref{eq1}) has been shown to support NI phases with $0<\theta<\pi/4$, AFTI phases with $\pi/4<\theta<3\pi/4$, and AFSOTI phases with $3\pi/4<\theta<5\pi/4$~\cite{PhysRevB.103.L041402}. In this paper, we start with the case $\theta=0.9\pi$, where the system resides in an AFSOTI with topological corner modes.}

\begin{figure}
\includegraphics[width=\linewidth]{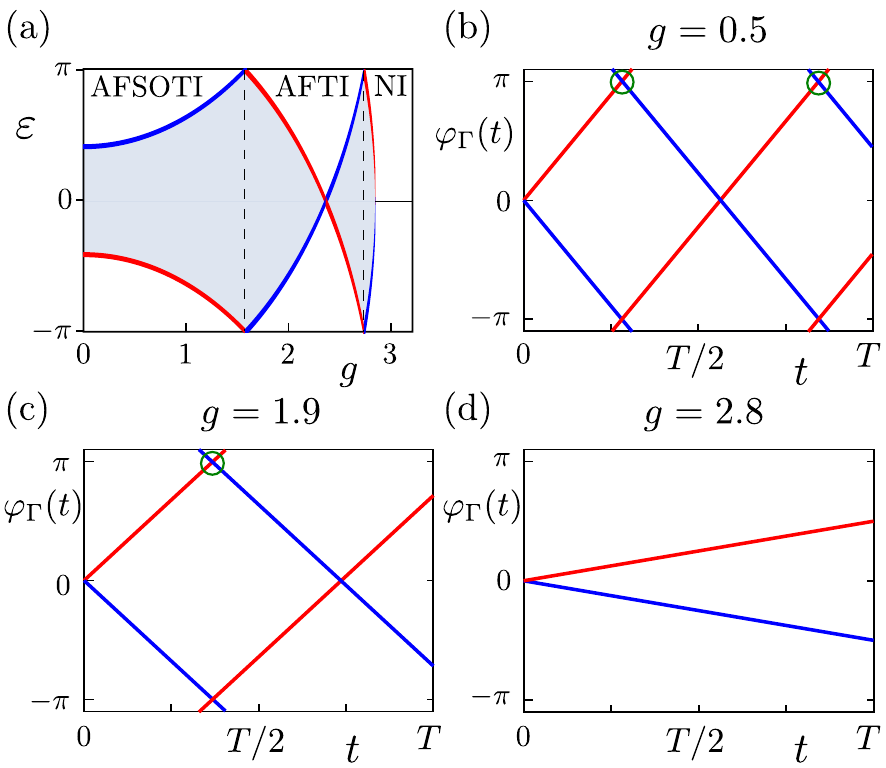}
\caption{Gain/loss-induced multiple topological transitions and their topological characterizations. (a) Bandwidth evolution as a function of $g$. (b)-(d) Topological characterization for (b) AFSOTI with $g=0.5$, (c) AFTI with $g=1.9$, and (d) NI with $g=2.8$. The coupling strength $\theta$ is set to $0.9\pi$.}
\label{TPT}
\end{figure}

\begin{figure*}
	\includegraphics[width=\linewidth]{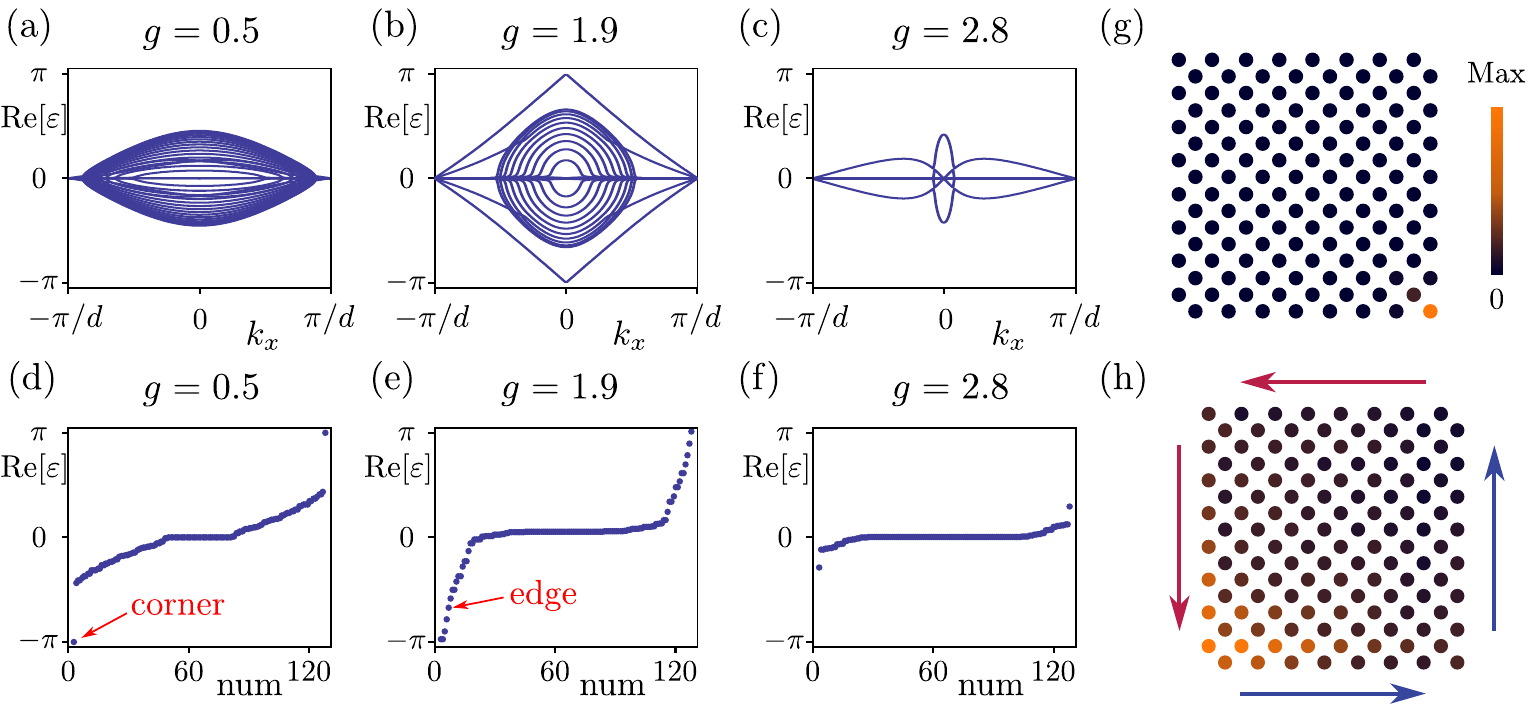}
	\caption{Quasienergy spectrum for the system with (a)-(c) $x-$PBC/$y-$OBC and (d)-(e) $x-$OBC/$y-$OBC. (a) and (d) are for the Floquet SOTI with $g=0.5$. (b) and (e) are for the anomalous Floquet topological insulator with $g=1.9$. (c) and (f) are for the normal insulator with $g=2.8$. (g) Eigenfield for one of the topological corner mode in (d). (h) Eigenfield for the topological edge mode in (e).}
	\label{spectrum}
\end{figure*}

\textcolor{black}{With the increase of gain and loss $g$, multiple topological transitions happen, as shown in Fig.~\ref{TPT}(a). The band gap at $\varepsilon=\pi$ closes and reopens at $g\approx1.56$, where the system goes from AFSOTI phases to AFTI phases. Increasing further the gain and loss, the bandwidth decreases to zero at $g\approx2.35$ and then increases. Another topological transition happens at $g\approx2.72$, where band gap closes and reopens, after which the AFTI phases switches to NI phases.}

\textcolor{black}{To prove that there are topological phase transitions, we provide topological descriptions for the states in our system. Since the bulk bands are all connected as a whole [Fig.~\ref{TPT}(a)], the topological states cannot be distinguished by usual topological invariants defined from the global property of eigenfunctions. In Hermitian systems, AFTI are described by singularities in the phase band, or equivalently by stable dynamical symmetry-inversion points when there are crystal symmetries~\cite{PhysRevB.104.L020302}. Here, we use dynamical phase bands at high symmetry points to characterize the topological states. Protected by PIS, the band inversion happens at high symmetry points,  which are the $\Gamma$ points in our model. Besides, due to the $\mathcal{PT}$ symmetry, the phase bands at $\Gamma$ points are real in the $\mathcal{PT}$-invariant region. The phase bands at $\Gamma$ points can be obtained from}
\begin{equation}\label{eq5}
U(\Gamma,t)\, |\psi(\Gamma,t)\rangle = e^{-i\varphi(\Gamma,t)} |\psi(\Gamma,t)\rangle,
\end{equation}
Here, $U(\Gamma,t)\equiv \mathfrak{T}\mathrm{exp}\big[\!-\!i\!\int_{0}^{t} H(\Gamma,\tau)\,d\tau\big]$ is the time evolution operator at $\Gamma$ points. For our model, the dynamical phase band can be found as
\begin{equation}\label{eq6}
  \varphi(\Gamma,t)=\pm\sqrt{\theta^2-g^2}t \mod(-\pi,\pi].
\end{equation}

\textcolor{black}{Figs.~\ref{TPT}(b)--\ref{TPT}(d) show the dynamical phase bands under different gain and loss. With small gain and loss (e.g., $g=0.5$), there are two singularities at $\varphi=\pi$. The system then belongs to AFSOTI phases. With larger gain and loss (e.g., $g=1.9$), there is only one singularity in the phase bands. The system then enters AFTI phases, which must be separated from the AFSOTI phases by a topological transition. Increasing further the gain and loss, a second topological transition would happen, after which there are no phase-band singularities (e.g., at $g=2.8$) and the system ends up in NI phases.}

\textcolor{black}{The existence of different topological phases can be confirmed by investigating Floquet bands under different boundary conditions. Figs.~\ref{spectrum}(a)-\ref{spectrum}(c) show the Floquet bands with periodic boundary condition along $x$ and open boundary condition along $y$ ($x-$PBC/$y-$OBC). With small gain and loss (e.g., $g=0.5$), the $\pi$ gap is empty [Fig.~\ref{spectrum}(a)]. When the gain and loss is large enough (e.g., $g=1.9$), a first topological transition has happened at $g\approx1.56$ and there are gapless edge modes in $\pi$ gap [Fig.~\ref{spectrum}(b)]. Increasing further the gain and loss (e.g., to $g=2.8$), a second topological transition has happened at $g\approx2.72$ and the $\pi$ gap becomes empty again [Fig.~\ref{spectrum}(c)]. The presence of topological corner and edge modes can be further confirmed by studying the quasienergy bands under open boundary conditions along both $x$ and $y$ axes ($x-$OBC/$y-$OBC), with results shown in Figs.~\ref{spectrum}(d)-\ref{spectrum}(f). We notice that there are topological corner modes in band gap for $g=0.5$ [Fig.~\ref{spectrum}(d)], and topological edge modes in band gap for $g=1.9$ [Fig.~\ref{spectrum}(e)]. These results are consistent with previous topological descriptions in terms of dynamical phase bands.}

\begin{figure*}
	\includegraphics[width=\linewidth]{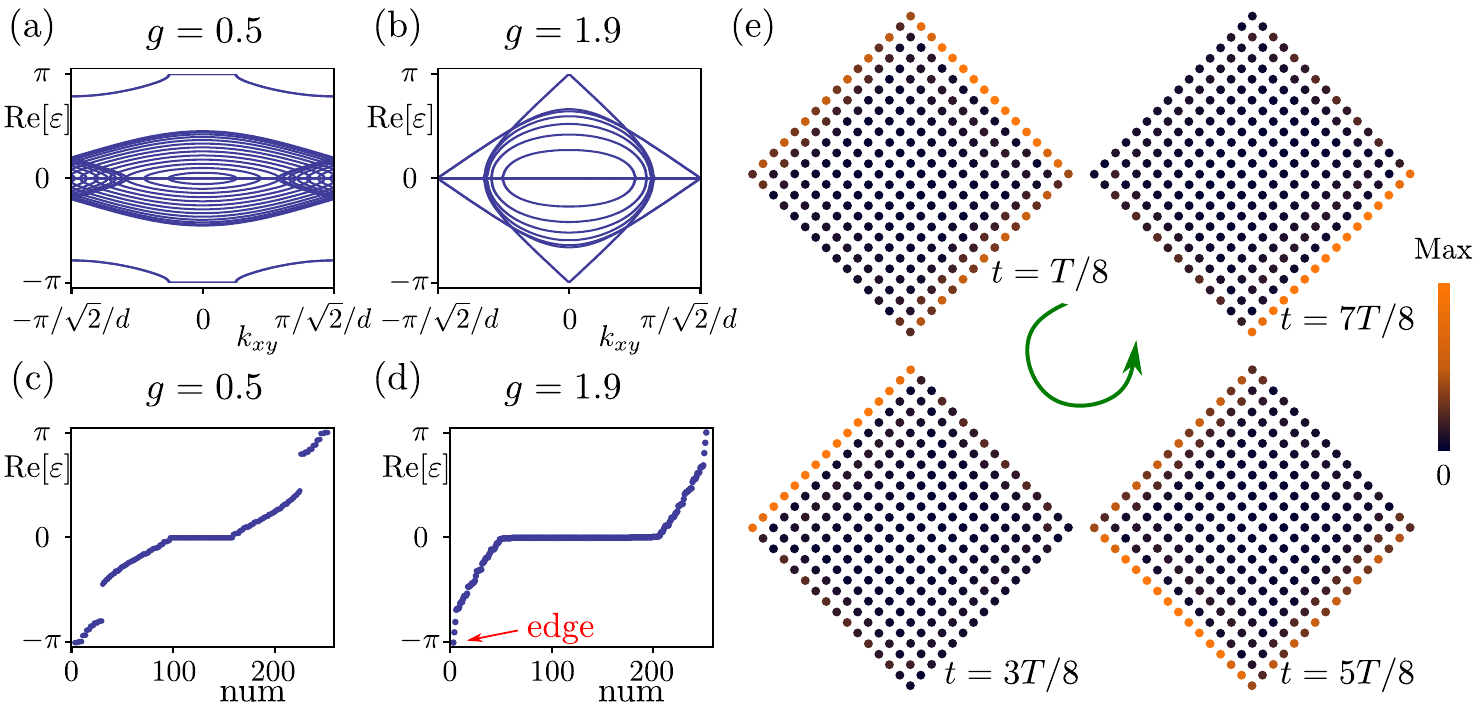}
	\caption{Quasienergy spectrum of the system with $xy-$PBC/$x\bar{y}-$OBC for (a), (b) and $xy-$OBC/$x\bar{y}-$OBC for (c), (d). (a) and (c) are for the SOTI with $g=0.5$. (b) and (d) are for the anomalous Floquet topological insulator with $g=1.9$. (e) Eigenfield for the topological edge state in (d) at different time slices.}
	\label{spectrumxy}
\end{figure*}

\textcolor{black}{We next investigate the topological boundary modes of the different phases in our system. The spatial distribution of one of corner modes for AFSOTI phase [Fig.~\ref{spectrum}(d)] is shown in Fig.~\ref{spectrum}(g). We notice that the state is localized at the right-down corner (the other one localized at up-let corner is now shown). Compared with the case in the corresponding Hermitian system, the only difference is that here the quasienergy is complex. The topological edge modes of the AFTI phase is localized at the left-down corner, as shown one example in Fig.~\ref{spectrum}(h), which is dramatically different from the extended topological boundary modes in the corresponding Hermitian system. Here, the localization originates from the combination of topological effects with gain and loss, which yields the hybrid skin-topological effect. The localization of topological edge states can be phenomenologically understood by the state accumulation shown in Fig.~\ref{spectrum}(h), where the topological edge state at the up-left (right-down) edge has gain (loss). Both of them are accumulated at the left-down corner. For $1.56\lesssim g\lesssim2.72$, the system simultaneously possesses anomalous Floquet topological insulator phases and hybrid skin-topological modes.}

\subsection{Geometry-dependent effect}\label{sec:GdE}

With small gain and loss $g\lesssim1.56$, the system resides in an AFSOTI phase. With larger gain and loss $1.56\lesssim g\lesssim2.72$, the system exhibits the hybrid skin-topological effect, which is a kind of second-order non-Hermitian skin effect. These second-order effects are usually protected by crystalline symmetries and the associated topological boundary states are geometry-dependent. It is then interesting to study the spectrum and eigenfield of the system with different boundary geometries.

In Fig.~\ref{spectrumxy}, we present the spectrum of the system in the geometry of a square at $45^\circ$ to the horizontal line. Figs.~\ref{spectrumxy}(a) and \ref{spectrumxy}(b) present the Floquet spectrum with PBC along $xy$ and OBC along $x\bar{y}$ ($xy-$PBC/$x\bar{y}-$OBC), which can be used to check topological edge states. With small gain and loss (e.g., $g=0.5$), there are counter-propagating edge states [Fig.~\ref{spectrumxy}(a)] for the AFSOTI, which is also a weak topological insulator. With larger gain and loss (e.g., $g=1.9$), there are gapless topological edge states [Fig.~\ref{spectrumxy}(b)]. Along the edge, the numbers of gain and loss sites are equal, so that the quasienergies of topological edge states are real. Figs.~\ref{spectrumxy}(c) and ~\ref{spectrumxy}(d) study the spectrum with OBC along both $xy$ and $x\bar{y}$ directions ($xy-$OBC/$x\bar{y}-$OBC). There are topological edge states isolated from the bulk states for $g=0.5$. Meanwhile, the edge states for the case with $g=1.9$ are gapless.

\begin{figure}
	\includegraphics[width=\linewidth]{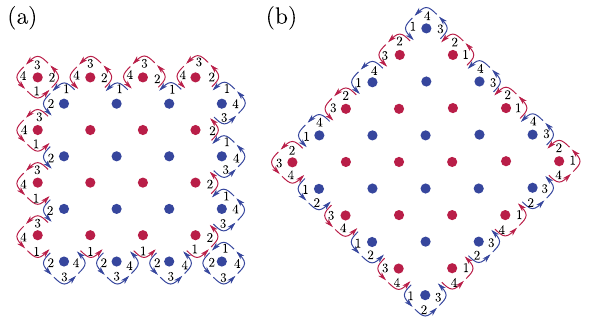}
	\caption{Evolution of topological edge modes in coupled ring resonators. (a) denotes the corner skin mode in Fig.~\ref{spectrum}(h). (b) denotes the hybrid localized-delocalized mode in Fig.~\ref{spectrumxy}(e).}
	\label{dynamical}
\end{figure}

\begin{figure*}
	\includegraphics[width=\linewidth]{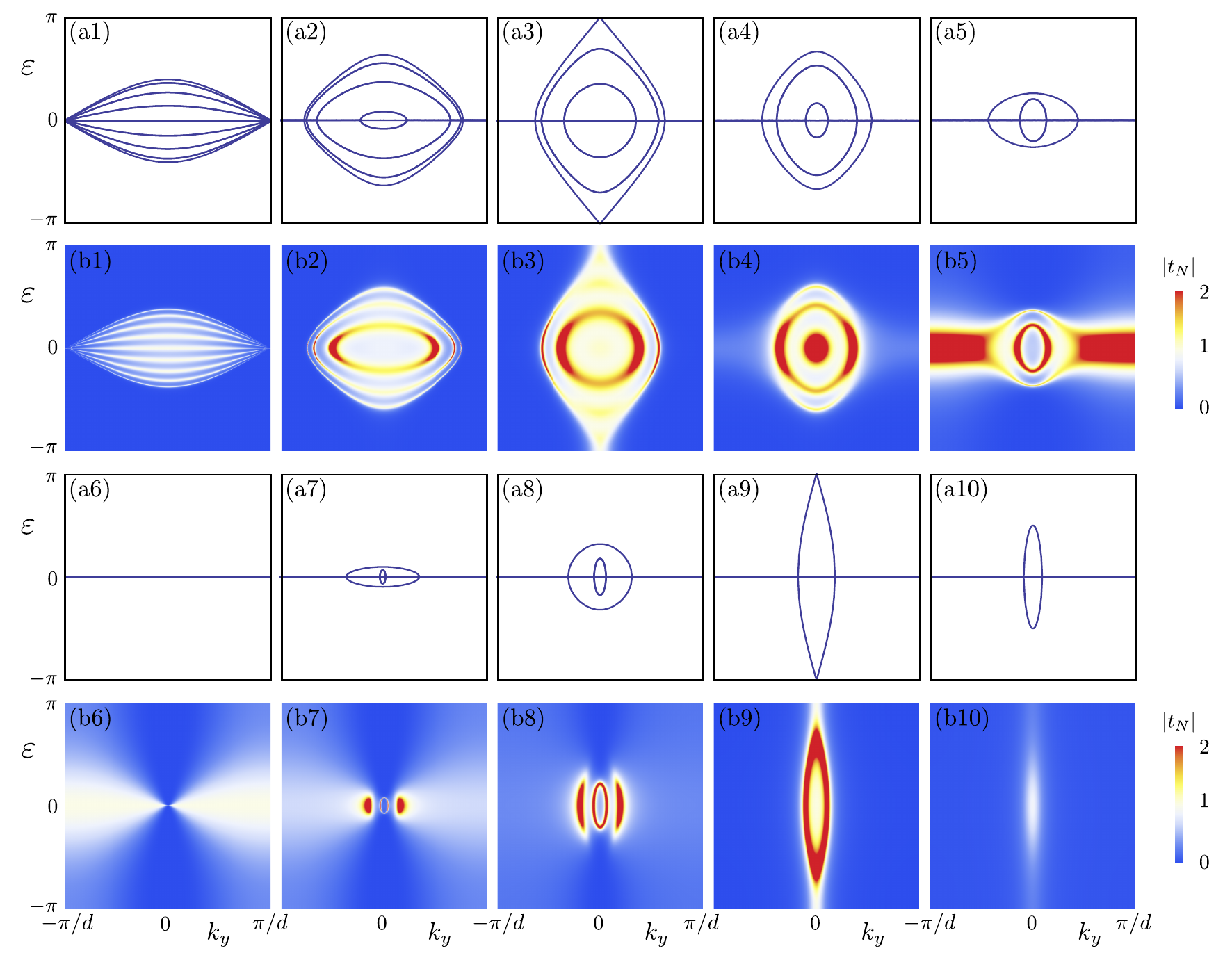}
	\caption{Quasienergy spectrum and transmission coefficients for the Floquet lattice. (a) Quasienergy spectrum with $x-$PBC/$y-$PBC. Eight unit cells are chosen along the $x$-axis. (b) Transmission coefficients $|t_N|$ as a function of $\varepsilon$ and $k_y$. Nine unit cells are chosen along the $x$-axis. The gain/loss magnitudes are $g=0$ for (a1), (b1), $g=1$ for (a2), (b2), $g\approx1.56$ for (a3), (b3), $g=1.9$ for (a4), (b4), $g=2.2$ for (a5), (b5), $g\approx2.35$ for (a6), (b6), $g=2.4$ for (a7), (b7), $g=2.5$ for (a8), (b8), $g\approx2.716$ for (a9), (b9), and $g=2.8$ for (a10), (b10).}
	\label{Transmission}
\end{figure*}

The eigenfield for the topological edge state in Fig.~\ref{spectrumxy}(d) is striking. As the quasienergy of the topological edge state with $xy-$PBC/$x\bar{y}-$OBC is real, there is no non-Hermitian skin effect for this edge states. However, the topological edge state is still localized at certain instant of time. As shown in Fig.~\ref{spectrumxy}(e), the topological edge state is localized at different edges at different evolution time within a driving period. Such a state is different from the topological chiral edge states in Hermitian systems, which are extended along the edge. And it is also different from the corner skin mode in Fig.~\ref{spectrum}(h), which is localized at one corner. Here, the topological edge state is localized at different time slices but traversing all edges within one driving period. This is an intriguing dynamical and topological phenomena and we term it a hybrid localized-delocalized mode.

\textcolor{black}{Such geometry-dependent effects can be understood by the evolution of topological edge states in coupled ring resonators array. For simplicity, we consider a typical case with $\theta=\pi/2$, in which the edge state evolution can be demonstrated graphically as shown in Fig.~\ref{dynamical}. Fig.~\ref{dynamical}(a) shows the result for a horizontal square geometry, and Fig.~\ref{dynamical}(b) shows the result for a square at $45^\circ$ to the horizontal line. In one driving period, the edge state evolves following the path marked by $1$, $2$, $3$ and $4$. We then consider adding gain and loss to the two sublattices. For the horizontal square in Fig.~\ref{dynamical}(a), the up and left edge states are dominated by gain sites, while the right and lower edge states are dominated by loss sites, forcing the eigenfields of topological edge states to accumulate at the left-down corner. For different time slices, the eigenfield distribution vary slightly. But overall, the fields are localized at the left-down corner. The situation is different for the square at $45^\circ$ to the horizontal line in Fig.~\ref{dynamical}(b). We notice that the numbers of gain and loss sites are equal at different edges, so the topological edge state is extended along the whole edge. However, at different time slices, the gain and loss experienced by each edge are different. In step 1, the right-up and right-down edges are dominated by gain, while the left-up and left-down edges are dominated by loss. The field is then localized at the right-up and right-down edges. In step 2, the right-up and left-up edges are dominated by gain, and the field is localized at the right-up and left-up edges. Similarly, we observe that the field is localized at the left-up and left-down edges (the left-down and right-down edges) in step 3 (4).}

\section{Spectral singularity}\label{sec:SS}

As mentioned in Sec.~\ref{sec:Mod}, the time-dependent Hamiltonian is equivalent to the coupled ring resonators array in Fig.~\ref{model}(b). So we can naturally study the transport properties of the system by the transfer matrix method. The transport porperties are hopefully reflective of some properties of the Floquet bands studied in the previous sections. In particular, the transmission coefficient of the system can be obtained from 
\begin{equation}\label{eq7}
  t_N=\frac{1}{T_{22}^{N}}.
\end{equation}
The reflection coefficient of the system is given by
\begin{equation}\label{eq8}
	r_N=-\frac{T_{21}^{N}}{T_{22}^{N}}.
\end{equation}
Here, $T_{22}^{N}$ is the second row and second column element of matrix $T^N$. $T_{21}^{N}$ is the second row and first column element of matrix $T^N$. $T^N$ is the transfer matrix for $N$ unit cells. It is obtained by taking the $N$th power of the matrix $M'M$, i.e.,
\begin{equation}\label{eq9}
 T^N=(M'M)^N.
\end{equation}

\textcolor{black}{Figs.~\ref{Transmission}(a1)--\ref{Transmission}(a10) show the bulk dispersion of Floquet bands vs $k_y$. Along the $x$-axis, we also choose the PBC with eight unit cells. Figs.~\ref{Transmission}(b1)--\ref{Transmission}(b10) show the transmission coefficients as functions of the quasienergy $\varepsilon$ and quasimomentum $k_y$. In our calculations, we set $N=5$. From the band structure and transmission coefficients, we find that that Floquet band gap close at $\pm \pi$ and hence  topological phase transitions happen at $g\approx1.56$ [Figs.~\ref{Transmission}(a3) and \ref{Transmission}(b3)] and $g\approx2.716$ [Figs.~\ref{Transmission}(a9) and \ref{Transmission}(b9)]. These results are consistent with our previous analysis. As such, in general the transmission properties will be able to detect various topological phase transitions induced by gain and loss. }

\textcolor{black}{Comparing the results of band-structure calculations and transmission coefficients, as expected from above,  they are mostly consistent with each other. That is, the dispersive bulk bands lead to nonzero transmission coefficients, whereas the flat bands and band gaps generally yield zero transmissions. Nevertheless, there are remarkable exceptions in certain cases. One obvious example is shown in Figs.~\ref{Transmission}(a5) and \ref{Transmission}(b5) ($g=2.2$), where the transmission coefficient is very large even though the Floquet bands are partially flat. Another example is given in  Figs.~\ref{Transmission}(a6) and \ref{Transmission}(b6) ($g\approx2.35$), where the transmission coefficient is unity even though the entire Floquet bands are flat. These anomalous transmission properties can be explained by the so-called spectral singularity or unit cell resonance \cite{Singularity2009}.} 

\begin{figure}
	\includegraphics[width=\linewidth]{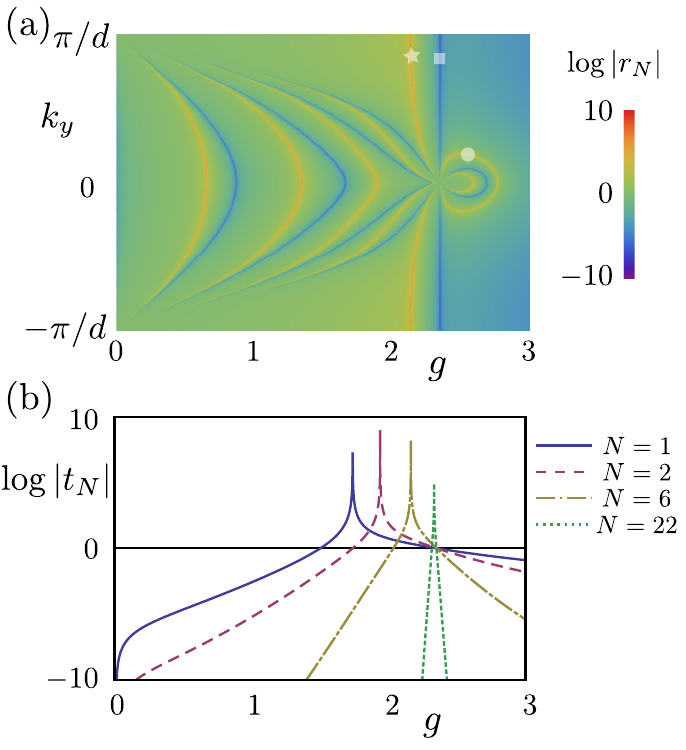}
	\caption{Spectral singularity. (a) Logarithm of the reflection coefficient $\log|r_N|$ for $N=5$ as a function of $k_y$ and $g$ at $\varepsilon=0$. (b) Logarithm of the transmission coefficient $\log|t_N|$ with fixed $k_y=\pi/d$ as a function of $g$. Solid, dashed, dash-dotted and dotted lines represent the results for $1$, $2$, $6$ and $22$ unit cells.}
	\label{singularity}
\end{figure}

\textcolor{black}{Let us now focus on the spectral singularity of the system. Spectral singularities are special points that spoil the completeness of eigenfunctions of certain non-Hermitian operators \cite{Singularity2009}. It has been shown to happen at $T_{22}^{N}=0$ for complex scattering potentials, where the reflection and transmission coefficients tend to infinity. Fig.~\ref{singularity}(a) shows the logarithm of reflection coefficient $\log|r_N|$ as a function of $k_y$ and $g$ at $\varepsilon=0$. We notice that there are several red lines corresponding to infinite reflection coefficients. The left four lines are related to multiple scatterings between different unit cells. The line marked by star is mainly determined by the property of unit cells. Different from other lines, here the spectral singularity exists at $k_y=\pi/d$ with $g\approx2.2$. It hence explains the large transmission coefficient at $k_y=\pi/d$ and $\varepsilon=0$ in Fig.~\ref{Transmission}(b5), in the presence of a flat quasienergy band in that regime. Besides, the positions of singularity points are size-dependent. Fig.~\ref{singularity}(b) shows the logarithm of transmission coefficient $\log|t_N|$  as a function of $g$ with a fixed $k_y=\pi/d$ for different numbers of unit cells. We find that the positions of singularity points shift to right with the increase of the unit-cell numbers. With a large number of unit cells, the positions of singularity points tend to coincide with the locations of unit cell resonance at $|t_N|=1$.}

From Fig.~\ref{singularity}(a), we observe another line marked by square, which corresponds to vanishing reflection coefficients. This line well explains the unit transmission at $\varepsilon=0$ in Fig.~\ref{Transmission}(b6), although the Floquet band is completely flat.

\section{Conclusion}\label{sec:Sum}

\textcolor{black}{In this work, we revealed two unique phenomena in Floquet systems induced by non-Hermitian gain and loss, i.e., the multiple topological transition and the spectral singularity.  Because of the multiple Floquet topological phase transitions,  we discover Floquet hybrid skin topological modes by adding appropriate gain and loss to anomalous Floquet higher-order topological insulators. Interestingly, the profile of the associated Floquet hybrid skin topological modes is found to depend on the geometry of the lattice. Though in general we see a direct link between Floquet band features and transmission properties,  we also discover that significant wave transmissions could appear even at flat bands due to the physics of spectral singularity. This represents another interplay between periodic driving and gain/loss effects. These intriguing properties might be observable in optical and acoustic experiments. The Floquet model proposed in this work can be realized in coupled waveguide system or coupled ring resonator array system and the non-Hermitian gain/loss can be replaced by loss difference~\cite{NatCommun7.13368,NatCommun8.13756,NatCommun13.11,Nature496.196,NM18}.}

\section{Acknowledgement}
W. Z. acknowledges support from the Start up Funding from Ocean University of China.
\textcolor{black}{L.~Z. is supported by the Fundamental Research Funds for the Central Universities (Grant No.~202364008), the National Natural Science Foundation of China (Grants No.~12275260, No.~12047503, and No.~11905211), and the Young Talents Project of Ocean University of China.}
L. L. is supported by National Natural Science Foundation of China (Grant No. 12104519) and the Guangdong Project (Grant No. 2021QN02X073). J. G. acknowledges support by the National
Research Foundation, Singapore and A*STAR under its
CQT Bridging Grant.

\end{document}